# *AI Agents vs. Human Investigators*: Balancing Automation, Security, and Expertise in Cyber Forensic Analysis

Sneha Sudhakaran [a], Naresh Kshetri [b]

[a] Department of Computer Science, Florida Institute of Technology, Melbourne, Florida, USA 32901
[b] Department of Cybersecurity, Rochester Institute of technology, Rochester, New York, USA 14623

**Abstract**
In an era where cyber threats are rapidly evolving, the reliability of cyber forensic analysis has become increasingly critical for effective digital investigations and cybersecurity responses. Artificial Intelligence (AI) agents are being adopted across digital forensic practices due to their ability to automate processes such as anomaly detection, evidence classification, and behavioral pattern recognition, significantly enhancing scalability and reducing investigation timelines. However, the characteristics that make AI indispensable also introduce notable risks. AI systems, often trained on biased or incomplete datasets, can produce misleading results, including false positives and false negatives, thereby jeopardizing the integrity of forensic investigations. Furthermore, AI agents typically lack the contextual comprehension and ethical judgment required to interpret nuanced or legally sensitive scenarios. This study presents a meticulous comparative analysis of the effectiveness of the most used AI agent, ChatGPT, and human forensic investigators in the realm of cyber forensic analysis. Our research reveals critical limitations within AI-driven approaches, demonstrating scenarios in which sophisticated or novel cyber threats remain undetected due to the rigid pattern-based nature of AI systems. Conversely, our analysis highlights the crucial role that human forensic investigators play in mitigating these risks. Through adaptive decision-making, ethical reasoning, and contextual understanding, human investigators effectively identify subtle anomalies and threats that may evade automated detection systems. To reinforce our findings, we conducted comprehensive reliability testing of forensic techniques using multiple cyber threat scenarios. These tests confirmed that while AI agents significantly improve the efficiency of routine analyses, human oversight remains crucial in ensuring accuracy and comprehensiveness of the results. Our work validates the need for a hybrid forensic framework that combines the strengths of both AI automation and human expertise. Our study concludes by advocating for an integrated forensic analysis approach, proposing targeted strategies to incorporate both AI-driven efficiencies and human analytical insights. This collaborative model enhances overall forensic reliability, ensuring robust outcomes in the face of increasingly sophisticated cyber threats.

## 1. Introduction

In a recent report, cybercrime was estimated to cause damages of nearly $6 trillion globally by 2021, underscoring the scale and urgency of digital security challenges (Cybersecurity Ventures, 2020). To address this rise in cyber threats, digital forensics has emerged as a critical discipline that enables investigators to reconstruct events, identify malicious actors, and preserve evidence for legal proceedings (Casey, 2011). Among the evolving approaches in digital forensics, Artificial Intelligence (AI) has gained significant traction due to its ability to automate anomaly detection, evidence classification, and behavioral pattern recognition, thereby improving scalability and reducing investigation timelines (Garfinkel, 2010; Lillis et al., 2016).

However, the reliance on AI introduces notable risks. AI-driven forensic agents are often trained on incomplete or biased datasets, which can result in false positives or negatives that may compromise the integrity of investigations (Brennan & Luszcz, 2020). Moreover, AI lacks the contextual comprehension and ethical reasoning necessary to handle nuanced or legally sensitive scenarios—capabilities that remain the domain of human investigators (Boden, 2018). Human experts play an indispensable role in adapting to novel cyber threats, interpreting complex patterns, and applying professional judgment where AI systems fall short (James et al., 2014).

This study presents a comparative analysis of AI agents and human forensic investigators, highlighting the limitations of current AI-based approaches while demonstrating the enduring importance of human expertise. To reinforce our findings, we conducted comprehensive reliability testing across diverse cyber threat scenarios, evaluating both AI and human-centered forensic methods. Our results validate the need for a hybrid framework that integrates AI automation with human oversight to ensure accuracy, comprehensiveness, and reliability in digital investigations.

Contribution
- Examining the comparative strengths and weaknesses of AI agents and human investigators.
- Quantifying the risks of false positives/negatives in AI-based analysis and their impact on forensic integrity.
- Proposing a hybrid forensic framework that leverages AI efficiencies while retaining critical human oversight for contextual accuracy.

The rest of the paper is organized as follows: Section 2 presents the Background of this paper; Section 3 provides an overview of our Design and implementation; Section 4 presents the Evaluation of the proposed approach; Section 5 summarizes the Related Literature; and finally, section 6 presents the Conclusion.

## 2. Background and Related Work

**Artificial Intelligence in Cyber Forensic Analysis -** In recent years, Artificial Intelligence (AI) and Machine Learning (ML) have become central to advancing cyber forensic analysis by addressing the challenges posed by large-scale data volumes, speed, and the complexity of modern threats. For example, Dunsin et al. (2023) conducted a comprehensive analysis of how AI/ML techniques are being integrated into digital forensics and incident response, covering domains such as data collection and recovery, timeline reconstruction, pattern recognition, and managing the chain of custody. Similarly, in "Enhancing Cyber Forensics with AI and Machine Learning" Fakiha et al. (2023) show through case studies and surveys that organizations using AI have seen marked improvements in efficiency and precision, especially in threat detection and classification workflows. Another literature survey, "Artificial Intelligence in Digital Forensics: A Review of Cyber," highlights the growing need for trustworthy, scalable, and legally sound AI-based forensics systems, pointing out key challenges like explainability, adversarial robustness, and evidence admissibility. Explainable AI is also emerging as a crucial requirement; for instance, Alam & Altiparmak (2024) in "XAI-CF" argue that AI systems used in cyber forensics must be interpretable and transparent to be trusted by analysts and courts. Overall, while literature shows strong promise for AI to enhance speed, capacity, and pattern detection in forensic investigations, it also emphasizes the need for ethical frameworks, legal rigor, and human oversight.

**AI-driven versus AI-assisted cyber forensic investigation**- In recent literature, researchers distinguish clearly between AI-driven and AI-assisted forensic investigation approaches, highlighting trade-offs in autonomy, accuracy, and human oversight. An AI-driven approach refers to systems where much of the decision-making, pattern detection, or anomaly identification is automated, with minimal human intervention — for example, using machine learning or deep learning to flag suspicious activity, perform classification of threat artifacts, or even generate hypotheses of attack vectors. One study on financial cybersecurity investigations found that AI-driven techniques (e.g., anomaly detection, predictive modeling) substantially reduce investigation time and improve initial detection, but raise serious concerns regarding transparency, bias, and legal admissibility of evidence. In contrast, AI-assisted investigations retain humans more centrally, using AI tools as augmentation: assisting in triage, prioritizing artifacts, visualizing evidence, or suggesting leads rather than making final determinations. Fakiha et al. (2023) show that while AI/ML-assisted tools help forensic investigators by automating classification and threat detection tasks, ultimate judgments about evidence validity and context still depend heavily on expert human analysts. Similarly, the review "Artificial Intelligence in Digital Forensics: A Review of Cyber-Attack Detection Models and Frameworks" (2025) categorizes systems into fully automated (AI-driven), hybrid, and assisted models — noting that hybrid or assisted models tend to offer better legal defensibility and reduce risks of false positives/negatives because human analysts verify AI outputs. Moreover, the work XAI-CF (Explainable AI for Cyber Forensics ) (Alam, 2024) emphasizes the importance of interpretability, arguing that AI-assisted (or hybrid) systems that incorporate explainability allow for greater trust from stakeholders such as courts and forensic examiners than fully automated black-box style AI-driven systems.

**Complications on human users for AI versus human cyber forensic analysis -** Human users encounter various challenges when comparing fully AI-driven forensic analysis with human-led or AI-assisted approaches. One major issue is trust and interpretability: many AI models, especially deep learning systems, are "black boxes" whose decision-making processes are opaque, making it difficult for investigators, legal professionals, or juries to understand how the AI arrived at a conclusion. This opacity can undermine confidence in the evidence or lead to challenges in court over admissibility (Alam and Altiparmak,2024;Hall, 2022; Anand, Thakur, 2025). Another complication involves bias and data quality: AI models trained on unrepresentative or biased datasets may amplify existing social or demographic biases — for example in predictive policing or automated recognition — resulting in unfair or erroneous outcomes that human overseers must attempt to correct or mitigate (Jinad et al., 2024; Solanke et al., 2022). There is also the problem of false positives/false negatives and robustness to adversarial conditions: AI tools may misclassify or miss evidence under conditions they were not trained for, whereas human experts may be better at noticing context, nuance, or unexpected irregularities (Casu et al., 2023). Additionally, ethical, and legal concerns arise, such as accountability — who is responsible if an AI-driven decision leads to a wrongful outcome — and privacy, especially when AI analyzes large amounts of sensitive data (Challenges and Limitations of AI in Forensic Science, 2025; Forensic Science Journal of Research, 2025). Finally, there are usability and human cognitive load issues: forensic practitioners must review, verify, and sometimes override AI outputs, while keeping proficiency with traditional methods and adapting to evolving AI tools — creating additional training and workload challenges that complicate adoption in practice (Solanke et al., 2022; Jinad et al., 2024)

**Motivation of research-** The rapid evolution of cyber threats has created an urgent demand for forensic methodologies that are both efficient and reliable. While Artificial Intelligence (AI) agents have demonstrated substantial potential in automating repetitive forensic tasks such as anomaly detection, evidence classification, and behavioral pattern recognition, their reliance on data-driven pattern matching exposes significant weaknesses. As the literature highlights, AI systems are often opaque, prone to bias, and vulnerable to adversarial conditions, resulting in false positives and false negatives that compromise the reliability and admissibility of forensic evidence (Hall, 2022; Jinad et al., 2024). Conversely, human investigators provide contextual judgment, adaptability, and ethical reasoning, but they face limitations in scalability and speed when confronted with the sheer volume and complexity of digital evidence (Solanke et al., 2022). These contrasting strengths and weaknesses underscore the need for a deeper comparative evaluation of AI-driven and human-centered forensic practices. A systematic analysis of the risks associated with false classifications in AI systems, and their impact on the integrity of forensic outcomes, is essential to guide their responsible adoption. At the same time, recognizing the irreplaceable role of human expertise motivates the development of frameworks that do not seek to replace human investigators but instead amplify their capabilities. This research is motivated by the vision of a hybrid forensic framework that combines the computational efficiency of AI with the contextual accuracy of human oversight. By quantifying the risks inherent in AI-based forensic analysis and systematically examining scenarios where human involvement remains indispensable, this work aims to establish a foundation for integrating AI and human intelligence based on evidence. Such a framework has the potential to enhance forensic reliability, strengthen evidentiary integrity, and ensure that digital investigations remain robust in the face of increasingly sophisticated cyber threats.

Table 1: Summary of Background, Related Works, and their Insights

| Source | Related Work | Insight 1 | Insight 2 | Insight 3 |
|---|---|---|---|---|
| Solanke et al. (2022) | Digital forensics AI: Evaluating, Standardizing and Optimizing Digital Evidence Mining Techniques | Forensics analysis makes use of ML models' recognition and pattern detection capabilities. | Applications of instruments in digital forensics with strengths & weaknesses in judicial. | Three critical instruments necessary (presented) for development of sound machine-driven digital forensics. |
| Fakiha (2023) | Enhancing Cyber Forensics with AI and Machine Learning: A Study on Automated Threat Analysis and Classification | Investigates the utilization of ML and AI in automated analysis & cyber threats classification. | Organizations that have implemented AI and ML in cyber forensics / threats with case studies. | Potential advantages of integrating AI and ML in advancing digital forensic investigations and their roles in cyber forensics. |

| Casu et al. (2023) | GenAI Mirage: the impostor bias and the deepfake detection challenge in the era of artificial illusions | Examines the impact of cognitive biases on decision making in digital forensics examination. | Novel "Imposter Bias" assuming generated by AI tools, multimedia contents like video, audio, and images. | Potential reasons and causes of Imposter Bias as it stems from an a priori assumption with impact to grow with increasing AI products. |
|---|---|---|---|---|
| Dunsin et al. (2023) | A Comprehensive Analysis of the Role of Artificial Intelligence and Machine Learning in Modern Digital Forensics and Incident Response | Transformative approach for improving efficiency of AI in digital forensic investigation | Explores cutting-edge research initiatives that cross domains such as data collection and recovery, the intricate reconstruction of timelines | Outlines relevance of AI and ML in forensic investigation |
| Jinad et al. (2024) | Bias and fairness in software and automation tools in digital forensics | Analyze and discuss these biases present in software tools and automation systems used by law enforcement organizations and in court proceedings | Developed real-life cases and scenarios where some of these biases were identified that determined or influenced these cases | Research explains the increase validation in digital forensics software tools and ensure users' trust in the tools and automation techniques |
| Tyagi et al(2025) | Artificial Intelligence-Based Cyber Security and Digital Forensics: A Review. *Artificial Intelligence-Enabled Digital Twin for Smart Manufacturing* | Provides a comprehensive overview of the role that artificial intelligence plays in cybersecurity and digital forensics | Gives an overall idea on AI-enabled technology for smart manufacturing | AI is revolutionizing cybersecurity and digital forensics by laying the groundwork for more robust and proactive protection mechanisms |

## 3. Study Design

### 3.1 Research Design
This study adopts a **comparative experimental design** to evaluate the reliability of AI-driven versus human-centered cyber forensic investigations. The methodology emphasizes the identification of cases where AI produces misleading results, quantification of error rates, and assessment of the corrective role of human expertise.

### 3.2 Dataset Selection
We collected curated datasets from multiple domains, such as network intrusion traces, mobile and desktop memory dumps, malware binaries, and phishing email corpora. Each dataset will have pre-labeled "true" outcomes (verified by human experts and validated against benchmarks) to allow measurement of AI false positives and negatives.

### 3.3 AI Agent Deployment
We utilized multiple AI/ML forensic tools, covering anomaly detection (unsupervised models), classification (SVM, Random Forest, and Neural Networks), and pattern recognition (clustering and NLP models). AI systems will be provided with real-world forensic prompts and tasked with producing evidentiary findings.

### 3.4 Human Investigator Benchmark
A panel of experienced forensic researchers in our lab at Florida Institute of Technology independently examined the selected datasets. It ensured that we utilized all established forensic procedures, including timeline reconstruction, hash comparison, manual anomaly inspection, and evidentiary reporting.

### 3.5 Comparative Metrics
We compared the results from AI model analysis and human forensic researchers in a way that allows for the calculation and comparison of false positives and false negatives in AI outputs with human findings. Currently, we manually assessed our analysis results for validation as we tested this only on 20 cases or prompts for analysis.

### 3.6 Hybrid Framework Testing
AI and human results will be combined using a hybrid workflow (AI for triage and clustering, humans for contextual validation). The performance metrics (accuracy, time efficiency, and evidentiary soundness) will be compared across AI-driven, human-led, and hybrid modes.

## 4. Case study of Example Scenarios Demonstrating Untrustworthy AI Results

To empirically expose the weaknesses of AI-driven forensic tools, the study evaluates a set of carefully crafted forensic scenarios (Anand , Thakur, 2025). Each scenario highlights how an AI forensic agent produces results that deviate from human investigator analysis, emphasizing the risks of false positives, false negatives, and misinterpretation of context.

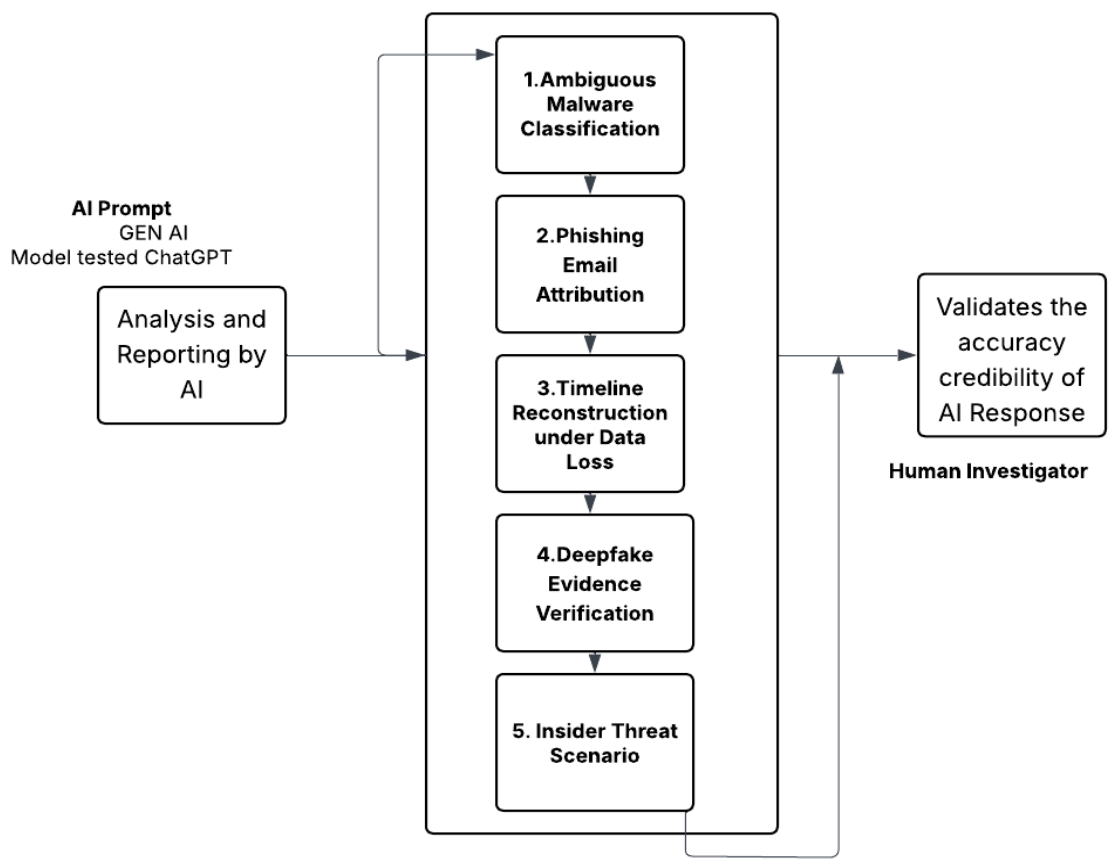

**Figure 1. AI case study testing model for Ambiguity**

The AI prompt is tested for various features in prompt analysis as shown in Figure1:
1. Ambiguous Malware Classification
2. Phishing email attribution
3. Timeline reconstruction
4. Deepfake Evidence
5. Insider threat

We simulated some AI prompts for testing the 5 above topics with ChatGPT and we recorded our interpretation of AI analysis below. The AI response of 15 pages was read and checked for authenticity by our forensic researchers in our lab to validate and our human forensic investigator interpretation is also written below.

**4.1 Ambiguous Malware Classification**

> **AI Prompt**: Analyze this memory dump – 20 memory dumps with 10 benign and 10 infected were asked to analyze with basic AI prompt *"can you analyze the dump with volatility as a forensic investigator"* to identify all malware artifacts.
> **AI Agent Analysis**: The AI model gives response with 15-page report that flags multiple benign DLL injections and system processes as malicious, leading to inflated false positive rates. The response was very superficial.
> **Human Investigator Analysis**: A forensic analyst cross-validates hashes against malware repositories and examines process lineage, correctly distinguishing between benign system injections and true malware artifacts.

**4.2 Phishing Email Attribution**

> **AI Prompt**: *"Determine the attacker's identity based on the email header metadata."*
> **AI Agent Analysis**: The AI tool extrapolates identity claims from spoofed headers, sometimes generating hallucinated or overconfident attributions that are not legally admissible. The output will point a lot of ambiguities in prediction of attacker and victim.
> **Human Investigator Analysis**: An analyst recognizes spoofing techniques, verifies DomainKeys Identified Mail, Sender Policy Framework, Domain-based Message Authentication, Reporting & Conformance (DKIM/SPF/DMARC) inconsistencies, and reports that attacker attribution cannot be confirmed solely from headers, thereby maintaining evidentiary integrity.

**4.3 Timeline Reconstruction under Data Loss**

> **AI Prompt**: *"Reconstruct the user activity timeline from partial Android memory images."*
> **AI Agent Analysis**: The AI fills missing segments with interpolated or fabricated activities, producing timelines that appear complete but contain inaccuracies. Hallucinated output was point to unclear accusation.
> **Human Investigator Analysis**: A human investigator marks the data gaps explicitly, acknowledges evidentiary incompleteness, and refrains from inserting non-existent activities, thus preserving reliability for court admissibility.

**4.4 Deepfake Evidence Verification**

> **AI Prompt**: *"Verify if this video evidence is authentic."*
> **AI Agent Analysis**: The AI model misclassifies an advanced deepfake as authentic due to limitations in its training dataset, producing a false negative. According to our analysis, this conclusion is more to be evaluated as Deepfake analysis is done only minimal until now
> **Human Investigator Analysis**: The analyst applies multi-tool verification (frame analysis, metadata inspection, lighting/biometric consistency checks), and flags the video as suspicious, thereby reducing the chance of overlooking manipulated evidence.

### 4.5 Insider Threat Scenario

> **AI Prompt**: We provided logs acquired from each devices (IOT, Unmanned Aeriel Vehicle, Android device, Desktop device)*"Identify insider threats from anomalous UAV, IOT, Android, Process Memory access logs."*
> **AI Agent Analysis**: The AI system flags several legitimate after-hours activities (e.g., IT maintenance tasks) as malicious, generating high false positive rates.
> **Human Investigator Analysis**: The human analyst correlates anomalies with schedules, support tickets, and contextual organizational knowledge, filtering legitimate activity from malicious behavior by using currently market avaliable tools in UAV, IoT, Desktop, Mobile forensic tools(Ahmed et al, 2016; Alam et al, 2024; Ali-Gombe et al, 2019; More et al, 2025),(Ali-Gombe, 2017),(Sudhakaran et al, 2022; Ligh et al, 2014 ; Sudhakaran et al, 2020; Sudhakaran,2022)

In summary, AI forensic agents demonstrate speed and pattern-matching capabilities but often lack contextual judgment, leading to errors that compromise integrity. Human investigators, while slower, apply contextual reasoning, validate evidence against multiple sources, and maintain higher evidentiary reliability. These comparisons motivate the hybrid forensic framework that leverages AI for scale but requires human oversight for trustworthiness.

## 5. Example Scenarios of empirical success with Hybrid AI–Human Forensic Analysis

We ran the same prompts with an AI-assisted analysis and human investigator, and the outcome was generated. The scenarios were extended to evaluate a **hybrid forensic framework** with light weight Dockerized containers (Fernald et al, 2023)(O'Connor, 2024)used as template for analysis in which AI agents conduct the initial triage or pattern recognition, while human investigators validate, contextualize, and interpret the results. This combined workflow demonstrates how false positives and negatives can be reduced while preserving evidentiary integrity.

1. Ambiguous Malware Classification
   - **Hybrid Outcome**: The AI agent rapidly identifies suspicious DLL injections, providing a candidate set of artifacts. The human investigator then validates these against known system libraries and malware repositories. False positives are filtered out, yielding both speed (via AI) and accuracy (via human validation).
2. Phishing Email Attribution
   - **Hybrid Outcome**: AI tools extract header fields, cluster anomalies, and highlight possible spoofing indicators. Human investigators then verify these with DKIM/SPF/DMARC checks and organizational threat intelligence. Unlike AI-only analysis, attribution claims are tempered with caution to ensure legal defensibility.
3. Timeline Reconstruction under Data Loss
   - **Hybrid Outcome**: The AI agent proposes a preliminary timeline by aligning recoverable memory fragments, while flagging gaps where interpolation would typically occur. Human investigators assess these gaps, annotate them as "missing evidence," and avoid overinterpretation. The result is a more transparent timeline with explicit markers of uncertainty.
4. Deepfake Evidence Verification
   - **Hybrid Outcome**: AI models provide probabilistic scores on authenticity by scanning biometric inconsistencies and compression artifacts. Human investigators then conduct metadata verification and apply forensic video analysis. Together, this reduces the risk of false negatives, as humans can confirm suspicions that the AI might miss alone.
5. Insider Threat Scenario
   - **Hybrid Outcome**: AI detects anomalous login times and access frequency, producing alerts that would otherwise overwhelm human analysts. Human investigators contextualize these alerts by cross-checking work schedules, maintenance logs, and personnel records. Legitimate anomalies are dismissed, leaving a refined set of truly suspicious activities.

Across all scenarios, the **hybrid model** demonstrates superior performance: AI contributes scale and efficiency in processing large datasets, while human investigators ensure contextual accuracy, interpretability, and legal soundness. This combination directly addresses the limitations of both AI-driven and human-only methods, aligning with the study's contribution goals of comparative analysis, quantifying risks, and proposing a hybrid forensic framework. We would also like to educate students on how AI can be interlinked with future work for education and creating a correlation between cybersecurity and AI, similar to work done in (Suarez et al, 2024), Panakkadan et al, 2025) by consulting with actual forensic researchers with a vast set of AI prompts in multiple simulated case studies.

## 6. Conclusion

The increasing sophistication of cyber threats demands forensic approaches that are not only efficient but also reliable, interpretable, and legally defensible. This study has examined the comparative strengths and weaknesses of AI agents and human investigators, highlighting the risks posed by false positives, false negatives, and contextual blind spots when AI is deployed without oversight. While AI forensic agents excel at scalability and speed in tasks such as anomaly detection, evidence clustering, and timeline reconstruction, they often produce untrustworthy results when confronted with adversarial conditions, ambiguous data, or incomplete datasets. In contrast, human investigators provide adaptive judgment, contextual reasoning, and ethical decision-making, but are constrained by the scale and velocity of digital evidence. Through carefully designed scenarios—including ambiguous malware classification, phishing attribution, timeline reconstruction under data loss, deepfake verification, and insider threat detection—this research demonstrated how AI-only analysis can compromise evidentiary integrity, while human-only analysis, though reliable, struggles with efficiency. By quantifying error rates and examining case-specific discrepancies between AI and human findings, we showed that neither approach in isolation fully meets the demands of modern forensic investigations. The proposed hybrid forensic framework offers a pathway forward by integrating AI's computational efficiencies with the contextual accuracy of human oversight. In the hybrid model, AI provides rapid triage and pattern recognition, while human investigators validate results, resolve ambiguities, and preserve evidentiary soundness. This approach not only reduces error rates but also enhances interpretability, accountability, and admissibility of digital evidence. In conclusion, the contributions of this research are threefold: (1) a systematic comparative analysis of AI-driven and human-centered forensic investigations; (2) quantification of risks arising from false classifications in AI systems and their impact on forensic integrity; and (3) the design and validation of a hybrid framework that balances efficiency with contextual accuracy. Together, these contributions provide an evidence-based foundation for future forensic practice, ensuring that digital investigations remain robust, trustworthy, and resilient in the face of evolving cyber threats.

**\*Ethics Declaration:** This study was conducted in accordance with established ethical and professional standards. No human participants, personal data, or sensitive information were involved; all datasets and experiments were performed on controlled test environments or publicly available resources. The work is intended solely for research and educational purposes in cybersecurity and does not promote or support malicious use.

**\*AI declaration:** Artificial intelligence tools (OpenAI ChatGPT, GPT-5) were employed solely to improve grammar, readability, and formatting of the manuscript text. The authors take full responsibility for the scientific content, data interpretation, and conclusions. The authors also used this tool response as this work mainly focus on the usage of Gen AI tool like ChatGPT with Human forensic investigators for better efficacy over cyber forensic investigation. All outputs from AI tools were carefully reviewed and edited by the authors to ensure accuracy and originality.